# Genealogical Information Search by Using Parent Bidirectional Breadth Algorithm and Rule Based Relationship

Sumitra Nuanmeesri

Department of Information Technology
King Mongkut's University of
Technology North Bangkok,
Bangkok, Thailand

Chanasak Baitiang

Department of Applied Science
King Mongkut's University of
Technology North Bangkok,
Bangkok, Thailand

Phayung Meesad

Department of Information Technology
King Mongkut's University of
Technology North Bangkok,
Bangkok, Thailand

*Abstract*—**Genealogical information is the best histories resources for culture study and cultural heritage. The genealogical research generally presents family information and depict tree diagram. This paper presents Parent Bidirectional Breadth Algorithm (PBBA) to find consanguine relationship between two persons. In addition, the paper utilizes rules based system in order to identify consanguine relationship. The study reveals that PBBA is fast to solve the genealogical information search problem and the Rule Based Relationship provides more benefits in blood relationship identification.**

## I. INTRODUCTION

Due to a rapid increase in the world population, therefore genealogical information typically contain large amount of families' information and relationship between members. Generally genealogical information of a house includes first name, last name, ID, date of birth, sex, father's ID, and mother's ID. It is not difficult to know weather or not two people in a house have the consanguine relationship. Example, when a daughter gets married and moves to live with her husband, she changes her last name. Her data information in her parent's house will be removed. Her information is in the new house (her husband house).When she has a daughter; her daughter might do the same thing, get married and move away. This moving of daughters occurs again and again. The problem arises when we want to know consanguine relationships of two people who have not live in the same house. There are two reasons that make this a hard problem. The first reason is that personal information is contained in only one house, the other reason is that more houses and more information is generated each year in the country. When investigating the consanguine relationships of two people, the problem knows what the short of relationship they have. It starts from the first person nobody knows the short of direction to link between the two people. This paper proposes to develop algorithms to find the consanguine relationship of two people.

WINGIS is a genealogical information system (GIS) developed at the University of Port Elizabeth (UPE) for capturing genealogical information [2]. The WINGIS is a Window-based system which allows users to enter the genealogical information in a database and search for people according to name, date of birth or ID number (unique to WINGIS). It currently contains information on more than 550,000 South Africans ranging from 1615 onwards. It is obvious that, searching and browsing information in such a large database are the big problems, especially when many people have the same or similar surnames.

Moreover, the visualization of large information spaces is also a common problem. The information needs to be displayed in a meaningful way in order to facilitate analysis and identification process. Family tree, however, are unique and do not correspond exactly to hierarchical data structures. Zoomable user interfaces, called ZoomTree [3,4], could facilitate the dynamic exploration and browsing of family trees within WINGIS [7]. It has been, therefore, used to address theses problems in similar situations, especially with hierarchical data [1, 8].

Yen and Chen explained the design theory and implementation ideas of GIS for Chinese genealogy. Method of using the XML technology to create the metadata of genealogy, maintaining relations among the individuals, and develop management and visualization utilities to represent pedigree information will be introduced [5, 6].

Expert System technology is applied in genealogy search. C Language Integrated Production System (CLIPS) is a productive development and delivery expert system which provides an environment for the construction of rule in the CLIPS library. PHLIPS is integrated software that provides PHP with a basic interface to a CLIPS library in order to help individuals to tract their member history and find relatives [9].

Ontology is a new alternative method to represent the consanguine relationship of family tree. Members in a family are represented by nodes and their consanguine relationships are represented by edges [10]. Ontology based knowledge is proposed prototype system for manage model for Chinese Genealogical Record of Very Important Persons in Nationalist Party and Communist Party of China [11].

Most review paper presented a family tree by using different presented technique, however, the papers started from only one person. In this paper, we show Parent Bidirectional Breadth Algorithm (PBBA) to find consanguine relationship between two persons by starting from individuals and try to link them together and Rule Based Relationship of identify name of consanguine relationship.

## II. PROBLEM FORMULATION

Everybody know their relative, such as uncle, aunt, niece, cousin and grandparent, however, they might not know all members of their families. Moreover, information in house does not include all family relationship information. Each house includes information of all people who live in the house. Each person's information consists of their first name, last name, gender, date of birth, father ID and Mother ID. Thus, other people in their family (for example uncle aunt grandparent) are not included in the information of their house. The problem rises when we want to investigate consanguine relation between two persons. There are three questions that are whether or not two persons are in the same family, what kind of the consanguine relationship do they have and how fast can one find the consanguine relationship. Those questions can be answered by using database that is present hereafter.

In this section, we represent problem related to the consanguine relationship. Given is a directed network G = (N, A). Node $i$ represents people. Directed arcs $(i, j)$ represent consanguine relationship from parent $(i)$ to son or daughter $(j)$. The name of relationship depends on gender of two people. If node $i$ and node $j$ are nodes at the starter and the end of a directed arc. Node $i$ is a parent and node $j$ is an offspring. If gender of node $i$ is male and the gender of node $j$ is female. The arc will represent relationship from a father to his daughters shown in Figure 1. When a person has father, mother, and three children, the network is depicted in Figure 2. Because there are million populations in a country, their network is very complicated, as it can be seen in Figure 3.

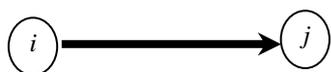

Figure 1 The relationship from node $i$ to node $j$.

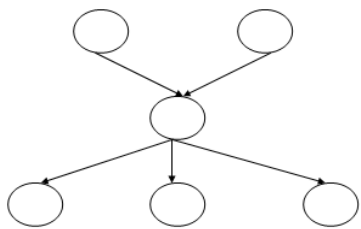

Figure 2 The direction from one node to the other nodes.

After we find an existence of the consanguine relationship between two persons, we should identify what kind of the relationship is there. We record the direction that is used to link the consanguine relation between them. In the next section, we describe detail of the methodology applied to find the consanguine relationship and rule to identify the relationship between two people.

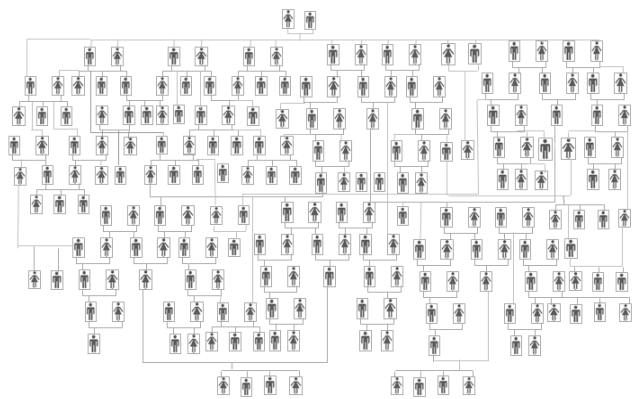

Figure 3 The complex relationship of population in a country.

## III. METHODOLOGY

In this section, we present PBBA to find consanguine relationship between two people and the Rule Based Relationship that is used to identify the relationship name. Parent Bidirectional Breadth Algorithm (PBBA) is developed by combine the Bidirectional approach and Breadth First Search and we define parent direction that is the only important direction for search. So, the algorithm is faster than the original Bidirectional approach and Breadth First Search for finding consanguine relationships.

Let $S$ be searching set that the program use to find the consanguine relationship, $P$ is a person node, $f_p$ is the father of person $P$ and $m_p$ is mother of person $P.x_p$ Represents kind of sex for person $P$, $A_p$ is age of person $P$, $B$ is binary variable. It equals one when a computer finds the consanguine relationship, it is zero if it is otherwise. The dept of searching is $L_{max}$. $L_c$ is current level for searching. In general, information of people includes father, mother, sex, and date of birth, and age of people can be calculate by using date of birth. Figure 4 shows procedure of Parent Bidirectional Breadth Algorithm.

After we discover that two people have the same blood relationship, we should identify the name of relationship too. That is the name of relationship depends on the level of generation, kind of sex, and age of person. Because of this, we can adapt rule according to countries.

In this paper, we present a rule based system using an English rule. First we classify people by two classes. The one class is descendant generation. That means offspring in family. The other one class is ancestor generation. That means the person who has previous generation of the lower class, for example, parent, grandparent, great-grandparent, etc.

The level is distance that count from one node to the intersection node (linking node). We compare between two people that want to know there relationship. The person node that has the level of generation more than the other person node is the descendant generation. Otherwise, the other person node is Ancestor generation. After that, the descendant generation uses name followed Figure 5 and the Ancestor generation uses name followed Figure 6.

| | |
|---|---|
| (1) | $S = \phi$ |
| (2) | $P_1 \rightarrow S_1$ |
| (3) | $P_2 \rightarrow S_2$ |
| (4) | $n = 3$ |
| (5) | $c = 1$ |
| (6) | **While** $B = 0 \wedge L_c < L_{max}$ |
| | { |
| (7) | **If** $S_c$ is intersection node of $P_1$ and $P_2$ **then** |
| (8) | Keep $R'_c$ |
| (9) | **else** |
| (10) | $f_c \rightarrow S_n$ |
| (11) | $m_c \rightarrow S_{n+1}$ |
| (12) | $R_n = c$ |
| (13) | $R_{n+1} = c$ |
| (14) | $n = n + 2$ |
| (15) | $c = c + 1$ |
| | } |
| (16) | **If** $B = 1$ |
| (17) | $S_{int}$ is intersection node |
| (18) | $R_{int}$ is node before $S_{int}$ |
| (19) | $S_c = R_{int}$ |
| (20) | **While** $(R_m = P_1) \vee (R_m = P_2)$ |
| | { |
| (21) | Find m that $R_m = S_c$ |
| (22) | $c = m$ |
| (23) | Collection the routing of family |
| | } |
| (24) | $S_c = R'_{int}$ |
| (25) | **While** $(R_m = P_1) \vee (R_m = P_2)$ |
| | { |
| (26) | Find m that $R_m = S_c$ |
| (27) | $c = m$ |
| (28) | Collection the routing of family |
| | } |
| (29) | **Else** |
| (30) | Two people are not in the same family. |

Figure 4 The procedure of Parent Bidirectional Breadth Algorithm.

In brief, the family relationship identification starts from finding the level of generation. Descendant generation uses rule from Figure 5 and ancestor generation uses Figure 6. From Figure 6, Line represents directly connection of two people. We use the line if ID of the parent is descendant generation which equals to the ID of ancestor generation. Form the family member who does not have the direct parent relationship, we use dash line to present. For example, if first and second person are different level of generation. First person is upper level and the second person is lower. The different level of two people is two levels. First person use Figure 6 and the second person use

Figure 5. That is the fist person might be grand father or grand mother if the connection between them is direct. On the other hand the first person might be a grand uncle or grand aunt if the connection is indirect. Form this example the computer compares data from the parent of the parent, so the second person has the same ID number with first person. The connection is direct; in addition, the computer checks what kind of data is stored about the first person. When the sex is a female, the first person is a grandmother. The second person uses the same technique with Figure 6. Kind of sex of the second person is male. This case the second person is a grandson.

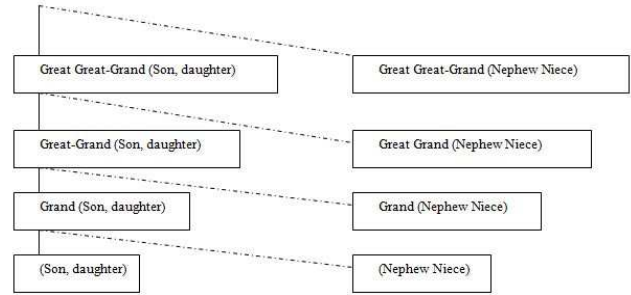

Figure 5 The name of relationship for descendant generation.

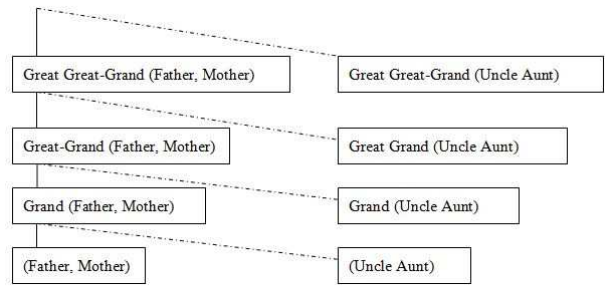

Figure 6 The name of relationship for ancestor generation.

## IV. EXPERIMENTAL RESULT

This section performance of PBBA is show by two ways. The first way is complexity. That is used to explain the difficult of the problem. If this algorithm is high performance, it should find the solution in a short computational time. We show performance of algorithm by proof mathematic. The second way is correctness of finding name of consanguine relationship. We test correctness by generating family data .Uniform distribution is use to generate data. The effectiveness of the PBBA is present by comparing with the effectiveness of BFS that is well known.

The effectiveness of BFS is presented. The algorithm starts from the first person node that wanted to link relationship to the second person node. From this first node, there are many nodes for choosing to move. They include of father node, mother node, and child nodes. That shows as follow Figure 7.

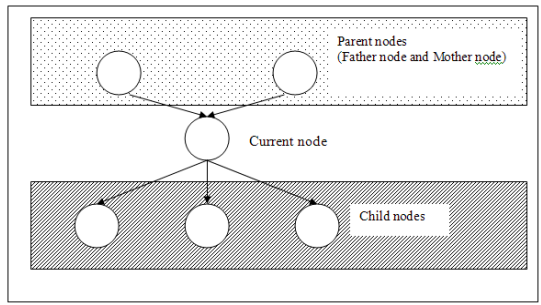

Figure 7 Relation between current node other nodes.

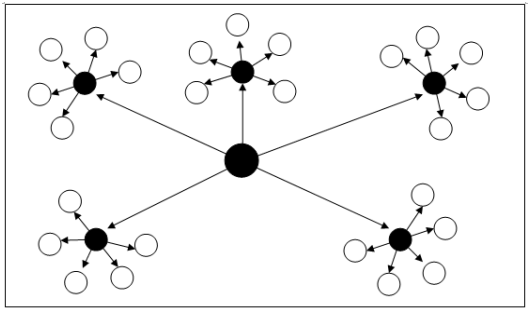

Figure 8 The alternative node for the second time search.

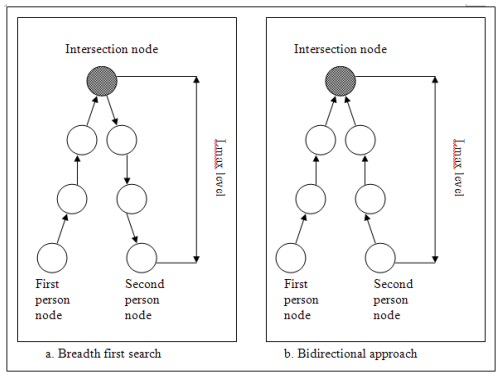

Figure 9 The directions and levels for search.

If $V$ is number of child node, there are $2+V$ alternative nodes (ways) for moving from current node to the other nodes. For the second time of searching, each node of $2+V$ has $2+V$ way to move. There are $(2+V)^L$ alternative nodes for moving. The alternative nodes are presented in Figure 8.

Trust, there are $(2+V)^L$ alternative nodes when the searching time is $L$. From the first person node to the second person node computer passes the node that is called Intersection node in Figure 9. Although the first person node and the second person node are at the same level in the family, the difference level between the level of the first person node and the intersection node is $L_{max}$ and the difference level between the level of the second person node and the intersection node is $L_{max}$. That is BFS must use level for linking relationship between the first and the second person nodes $2L_{max}$.

Total numbers of searching node are $\sum_{L=1}^{2L_{max}}(2+V)^L$ nodes. If one node spends one unit time, $\sum_{L=1}^{2L_{max}}(2+V)^L$ nodes spend $\sum_{L=1}^{2L_{max}}(2+V)^L$ time too.

PBBA performs in this problem with a shorter time. When the direction to search is parent node direction, the first person node has 2 ways for moving. Those are father node and mother node. In the next step, the father node and the mother node also have 2 nodes for moving. For $L_{max}$ steps, there are $2^{L_{max}}$ nodes for moving. For Bidirectional approach, the first person node and the second person node can search concurrently at the same time with the same behavior. The numbers of moving nodes from the first and second nodes are $2 \times 2^L$ nodes at $L_{max}$ steps. By the same reason, PBBA use $2\sum_{L=1}^{L_{max}}2^L$ time to find the solution.

If the numbers of moving nodes of two algorithms are compared, each node of breath first search algorithm has $(2+V)$ moving nodes and each node of PBBA has 2 moving nodes. In the fact, the bread first search cannot reduce nodes by searching with parent node direction because it could not find the solution. On the other hand, Bidirectional approach can reduce because it also finds the solution. Absolutely, PBBA is faster than the BFS in consanguine relationship between two persons.

When we compare the computational time of those two algorithms, we meet that $\sum_{L=1}^{2L_{max}}(2+V)^L$ times are more $2\sum_{L=1}^{L_{max}}2^L$ than times. So, PBBA is higher performance than BFS for the consanguine relationship problem.

Table 1 shows the comparison data of the BFS, Bidirectional approach and PBBA. The first column shows level of searching. The second column contains numbers of nodes that are used if the consanguine relationship, which can be found between $L$ levels. The third and the fourth columns contain the number of nodes that are used when the solution can be found between $L$ levels. Figure 10 shows the trend of used nodes for the three algorithms.

Figure10 reveals that BFS and Bidirectional approach uses many nodes to find the solution and PBBA used a low amount of nodes to find the solution. The fact that a low amount of nodes spent a short time to calculate the results shows that the PBBA is a best algorithm to solve the consanguine relationship problem.

We present correctness of the PBBA by testing algorithm with family data. First, we generate one family data. After that, we use the algorithm to find consanguine relationship from two people from the data. If the results are correct, we add the new family data to the old data. So, the new database includes two families. Then we run the algorithm to test correctness again. If the correctness satisfies, this process runs again until there are five families in the data.

TABLE I. THE DIRECTIONS AND LEVELS FOR SEARCH.

| Level | BFS | Bidirectional approach | PBBA |
|---|---|---|---|
| 1 | 30 | 10 | 4 |
| 2 | 780 | 60 | 12 |
| 3 | 19530 | 310 | 28 |
| 4 | 488280 | 1560 | 60 |
| 5 | 12207030 | 7810 | 124 |
| 6 | 305175780 | 39060 | 252 |
| 7 | 7629394530 | 195310 | 508 |
| 8 | 4.76837E+12 | 976560 | 1020 |
| 9 | 4.76837E+12 | 4882810 | 2044 |
| 10 | 1.19209E+14 | 24414060 | 4092 |

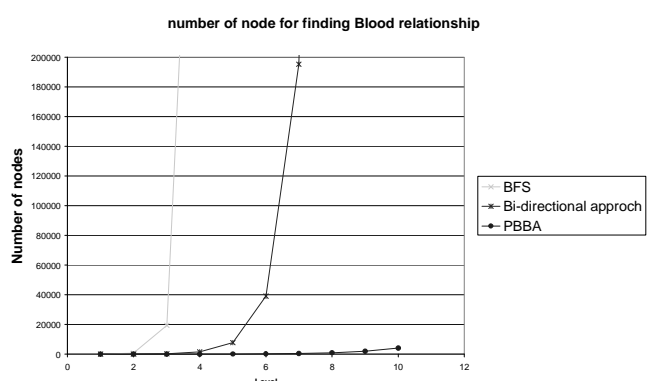

Figure 10 Comparison the effectives of three algorithms by the increasing of node number.

We start from one man, the computer generates his wife. Absolutely the man and his wife should have the same last name. In addition, the computer generates the next generation. There are *V* numbers of children. Sex of children is randomly generated. Each offspring has fifty percent to be a man and fifty percent to be women. After that, all children must be married and they must have new offspring. If the offspring is male, he has the last name of his father. When offspring is women, her last name that is changed depends on her husband. Her husband's name and his last name are generated randomly. The family is created until limited number of generation. Final we random two people from this data and check consanguine relationship. If all solution is correct. We add new family by the same technique to database and random to check solution again. When all solutions are correct, we add a new family to database again. We check the correctness of the algorithm until there are five families including in database.

There are two errors for correctness test. The first one is computer identify that two people have the same family but they have not. In this case, we check by computer routing to link relation between the people. If the computer can link routing completely, two people are the same family. If computer cannot link the routing, two people are not the same family. The second error is if the computer identifies that they have the some consanguine relationship, but they do not have the same consanguine. In this case, we create one column to check family data. The people who are in the same family have the same value in the checking column. For example, the members of the first family have value one and the members of the second have value two. This column is not load to the memory because it use for checking the correctness by human. When computer identify that two people are not the same family, we can check the correctness by using this column.

TABLE II. THE EXAMPLE OF GENERATED PEOPLE DATA IN DATABASE.

| Number Of Generation | Number Of Offspring | The Couple for Experiment | Number of Family | Number of pupation | Correctness |
|---|---|---|---|---|---|
|  |  | 5 | 1 | 254 | 100% |
| 7 | 2 | 5 | 2 | 508 | 100% |
|  |  | 5 | 3 | 762 | 100% |
|  |  | 5 | 1 | 2186 | 100% |
| 7 | 3 | 5 | 2 | 4372 | 100% |
|  |  | 5 | 3 | 6558 | 100% |
|  |  | 5 | 1 | 10922 | 100% |
| 7 | 4 | 5 | 2 | 21844 | 100% |
|  |  | 5 | 3 | 32766 | 100% |
|  |  | 5 | 1 | 39062 | 100% |
| 7 | 5 | 5 | 2 | 78124 | 100% |
|  |  | 5 | 3 | 117186 | 100% |

Table 2 shows at first the program starts with 2 descendants per family, only 1 family is created then 5 pairs of each family member will be picked up randomly. Next program will search for their genealogical information. Second step program will create 2 families, each family has 2 descendants then 5 pairs of each family member will be picked up randomly and each pair will be looked up. Next step the all same old processes have to be done according to Table 2. Then test its accuracy 60 times. All tests have shown that even though there family name has changed the search of genealogical information with rule-base relationship is still valid. The last one is the module to find blood relationship between two people from the large population in a database. The study reveals the result is correct 100 percent.

## V. CONCLUSION AND FUTURE WORK

Parent Bidirectional Breadth Algorithm (PBBA) provides higher effective performances to solve the genealogical information search problem. This method is a beneficial technique to find the connection between two people due to PBBA being able to link relationship between two nodes and parent node only. The parent direction technique helps to minimize unnecessary directions, thus, the computational time of this algorithm is short. In addition, we present Rule Based Relationship for identifying name or type of relationship; we used English rule as an example because it is a good guideline. Readers can adapt it for their own countries. For the future research, we will develop a new algorithm to explore the consanguine relationship of the group of people. The new algorithm can cluster several people into a group regarding their family.

AUTHORS PROFILE

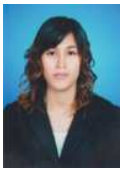

**Sumitra Nuanmeesir** received the B.S. degree (1[nd] Class Honor) in General Management, M.S. degree in Information Technology. She is currently Ph.D. Student in Information Technology. Her current research interests Information System, Algorithm and Application, Expert System, Database System.

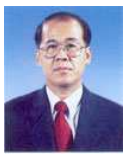

**Chanasak Baitiang** received the B.Ed. (2[nd] Class Honor), M.S. degree in Mathematics., and Ph.D. degree in Higher Education. His current research interests Discrete Mathematics, Graph Theory, Applied Linear Algebra, Model of Computation.

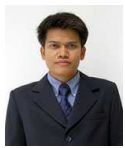

**Phayung Meesad** received the B.S., M.S., and Ph.D. degree in Electrical Engineering. His current research interests Fuzzy Systems and Neural Networks, Evolutionary Computation and Discrete Control Systems.